\documentclass{aa}
\usepackage{graphicx}

\begin{document}

\title{Relaxation times calculated from angular deflections}
\author{E. Athanassoula \and  Ch. L. Vozikis\thanks{\emph{Present
address:} Department of Physics, Section of Astro\-physics, Astronomy and
Mechanics,
Aristotle University of Thessaloniki, 54006 Thessaloniki, Greece}
\and  J. C. Lambert}
\institute{
Observatoire de Marseille,
2 place Le Verrier,
F-13248 Marseille Cedex 4, France}
\date{Received......, 2001 / Accepted......, 2001} 
\abstract{
In this paper we measure the two-body relaxation time from the angular
deflection of test particles launched in a rigid configuration of field
particles. We find that centrally concentrated configurations have
relaxation times that can be shorter than those of the corresponding
homogeneous 
distributions by an order of magnitude or more. For homogeneous
distributions we confirm that the relaxation time is proportional to
the number of particles. On the other hand centrally concentrated
configurations have a much shallower dependence, particularly for
small values of the softening. The relaxation time increases with the
inter-particle velocities and with softening. The latter dependence is
not very strong, of the order of a factor of two when the softening is
increased by an order of magnitude. Finally we show that relaxation
times are the same on GRAPE-3 and GRAPE-4, dedicated computer boards
with limited and high precision respectively.
}
\maketitle

\section{Introduction}
 
Collisionless $N$-body simulations are heavily used for studying the dynamical
evolution of galaxies, or systems of galaxies, and have so far been
remarkably successful in producing many interesting 
results. Yet proper care has to be taken to eliminate possible sources
of numerical errors which could, if present, lead
to erroneous results. One of the possible sources of errors
stems from the fact that the 
number of particles in a simulation is several orders of magnitude
less than the number of stars in a typical galaxy, or, in other words,
that the graininess in a computer realisation is much higher than that
of the galactic system it is meant to represent. This could lead to
errors since a particle moving through an $N$-body representation of a given
continuous system representing a galaxy is deflected from the orbit
it would have had in the corresponding smooth continuous medium, due to
two-body encounters.
This effect is known as two-body relaxation and the
characteristic time linked to it as two-body relaxation time
(hereafter $T_\mathrm{relax})$. 

The relaxation time will obviously increase with the number of particles
in the configuration, 
and tend to infinity as the number of particles tends to infinity, in
which limit the evolution of the system will follow the collisionless
Boltzmann equation. Thus the
relaxation time of simulations will be much shorter than that of
galaxies, which is much longer than the age of the universe. 
Relaxation leads to a loss of memory of the initial conditions and
an evolution of the system towards a state of higher entropy.
It is thus necessary to
have good estimates of the relaxation times of $N$-body simulations, 
since we can trust their 
results only for times considerably shorter than that. 

In the early times of $N$-body simulations, when the number of particles
used was of the order of a few hundred, the authors by necessity gave
estimates of relaxation times in order to enhance the credibility of
their results. Unfortunately in most cases only 
simple analytical estimates were used and the corresponding relaxation
times were found to be comfortably, although perhaps unrealistically, high. As
computers became faster, the number of particles used in simulations
was increased. Authors using several tens or hundreds
of thousands particles deemed unnecessary to include such simple
estimates of the relaxation times, since it was well known that the
simple analytical estimates would give reassuringly high relaxation
times. Nevertheless, it is not clear whether the simple
analytical estimates are in all cases sufficiently near the true
values. This could well be doubted since the simple analytical
estimates rely on a number of approximations, which are not in all cases
valid. 

Since different $N$-body methods may lead to different
relaxation rates, it is of interest to discuss relaxation times
when introducing a new method. It could thus have been feared that in
a tree code (Barnes \& Hut \cite{BarnesHut}) the relaxation time would not be
determined by the number 
of particles, but by the number of nodes, which would then act as
``super-particles''. Since the number of nodes is always much smaller
than the number of particles this would entail considerably shorter
relaxation times than direct summation with the same number of
particles, and thus constitute a major disadvantage of the tree
code. This fear was put to rest by Hernquist (\cite{hernq87}) who
showed that the relaxation for tree code 
calculations does not differ greatly from that obtained by direct
summation provided the tolerance parameter is less than 1.2. Similarly
Hernquist \& Barnes 
(\cite{hernq90}) compare relaxation rates in direct summation, tree and
spherical harmonic $N$-body codes, while Weinberg (\cite{weinberg})
introduces a modification of the orthogonal function potential solver
that minimises relaxation.

Several methods have been used to measure two-body
relaxation. Standish \& Aksnes (\cite{standish}), Lecar \&
Cruz-Conz\'alez (\cite{lecar}) and Hernquist (\cite{hernq87}) have
measured the angular deflection of test 
particles moving in a configuration of $N$ field particles. Although
this method has the disadvantage of not including collective effects,
it has the advantage that all the parameters can be changed
independently of each other, and that the results are easy to
interpret. Theis (\cite{theis}) performed semi-analytical calculations,
assuming a homogeneous medium and also
ignoring cumulative effects. The most widely used approach is to
monitor the energy conservation of individual particles in systems in
which, had it not been for the individual encounters, the individual
energies would have been conserved. This method, which includes collective
effects, is well suited for testing relaxation rates introduced by
different codes, but can only be used with systems in equilibrium. It has
been used e.g. by Hernquist \& Barnes (\cite{hernq90}), Hernquist \& Ostriker
(\cite{hernqOstr}), Huang, Dubinski \& Carlberg (\cite{huang}) 
and Weinberg (\cite{weinberg}). Theuns (\cite{theuns}) measured the
diffusion coefficients as a function of the energy in a direct
summation $N$-body simulation by studying the properties of the random
walk in energy space for particles of given energy and found very good
agreement with theoretically calculated diffusion coefficients.
Finally a number of studies (e.g. Farouki \& Salpeter \cite{farouki1},
\cite{farouki2}; Smith \cite{smith}) rely on a measurement of the mass
segregation, i.e. on the 
fact that, due to two-body encounters, high mass particles lose energy
and spiral towards the center, 
while light ones gain energy and move to larger radii. Thus the
configuration of the high mass particles contracts, while that of the
light particles expands, and from the rate at which this happens we
can calculate the two-body relaxation time.

In this paper we will calculate the relaxation times in a large number
of cases, using the first of the methods mentioned above, i.e. by
measuring deflection angles of individual trajectories of test
particles in a configuration of rigid field particles. We will
cover a much larger part of the parameter space than was done so far, and we
will also extend to larger number of particles. All the calculations
presented in this paper were 
made on the Marseille GRAPE-3, GRAPE-4 and GRAPE-5 systems. 
The Marseille GRAPE-3 systems have been described by
Athanassoula et al. (\cite{ABLM98}), while a general description of
the GRAPE-4 systems and their PCI interfaces has been given by Makino
et al. (\cite{makinoetal}) 
and Kawai et al. (\cite{kaw1}) respectively. A description of the
GRAPE-5 board can be found in Kawai et al. (\cite{kaw2}).
Opting for a GRAPE system restricts us to a single type of
softening, the standard Plummer softening, but has the big
advantage of allowing us to make a very large number of trials,
covering well the relatively large parameter space. Theis (\cite{theis}) 
compared the relaxation rates obtained with the
standard Plummer softening to those given by a spline (Hernquist \&
Katz \cite{hernqkatz}) and showed that the differences between the two
are only of the order of 20-40\%.

This paper is organised as follows: In
section~\ref{sec:analytical} we briefly summarise the
simple analytical estimates of the 
relaxation time. In section~\ref{sec:numerical} we describe
the numerical methods used in this paper and discuss the validity of
their approximations. Here we also introduce the mass models which
will be used throughout this paper. The values of the parameters to be
used, and in particular the values of the softening, are derived and
discussed in section~\ref{sec:free_param}. In section~\ref{sec:angle}
we give results for the 
relaxation time. We specifically discuss the effect of number of
particles, of the velocity and of the softening,
and compare results obtained with GRAPE-3 and GRAPE-4. We also give a
prediction for the relaxation time in an $N$-body simulation. We
summarise in section~\ref{sec:discussion}.

\section{Simple analytical estimates of the relaxation time}
\label{sec:analytical}

The relaxation time for a single star can be defined as the time
necessary for two body encounters to change its velocity, or energy, by
an amount of the same order as the initial velocity, or energy,
i.e. the time in which the memory of the initial values is lost. Thus
for the velocity we have

\begin{equation}
T_\mathrm{relax} = T_\mathrm{cross} \frac {\upsilon^2}{\Delta\upsilon_\perp^2},
\label{eq:Trelax1}
\end{equation}

\noindent
where $T_\mathrm{cross}$ is the crossing time of the system. 
Following e.g. Binney \& Tremaine (\cite{BT}) we can obtain a simple
order-of-magnitude estimate of the relaxation time. 
Let us
focus on the motion of a single star assuming that it
is moving on a straight line with a constant velocity ${\bf v}$ and that the
remaining stars have equal mass $m$ and are distributed uniformly in
space. We first consider that the  
star passes a single perturber star on its rectilinear trajectory. Then
the total change of its velocity component perpendicular to the trajectory is

\begin{equation}
\vert \delta\upsilon_\perp \vert \approx {{ 2 G m } \over {b \upsilon}},
\label{eq:single_dev_BT}
\end{equation}

\noindent
where $G$ is the gravitational constant and $b$ is the impact parameter,
i.e. the minimum distance between the 
two stars if there was no gravitational attraction. To find the total
change, due to all 
the particles, we integrate over all encounters and find

\begin{equation}
\Delta\upsilon_\perp^2 = \int_{b_\mathrm{min}}^{b_\mathrm{max}}
\delta\upsilon_\perp^2 
\approx \frac{ 8 G^2 m^2 N}{R^2 \upsilon^2} \ln (\frac {b_\mathrm{max}} {b_\mathrm{min}}),
\label{eq:total_dev_BT}
\end{equation}

\noindent
where $N$ is the number of stars, $m$ is their mass, $R$ is some
characteristic radius of the system 
and $b_\mathrm{min}$ and $b_\mathrm{max}$ are the minimum and
maximum values for the impact parameter. Using the same approximations 
and including a softening $\epsilon$, Huang, Dubinski \& Carlberg
(\cite{huang}) find 
\begin{eqnarray}
\Delta\upsilon_\perp^2  &=& \frac{ 4 G^2 m^2 N}{R^2 \upsilon^2} \nonumber\\
&&\left[ 
\frac{\epsilon^2}{b_\mathrm{max}^2 + \epsilon^2}
-\frac{\epsilon^2}{b_\mathrm{min}^2 + \epsilon^2}
+\ln\frac{b_\mathrm{max}^2 + \epsilon^2} 
{b_\mathrm{min}^2 + \epsilon^2} \right]. 
\label{eq:HDC}
\end{eqnarray}

A somewhat more elaborate derivation following the calculation of the
diffusion coefficients (e.g. Chandrasekhar \cite{chandra}, Spitzer
\& Hart \cite{spitzer2}, Spitzer \cite{spitzer1}, Binney \& Tremaine
\cite{BT}) gives 
\begin{equation}
T_\mathrm{relax} = {{F \upsilon^3} \over {G^2 m \rho~\ln ( {b_\mathrm{max}} /
{b_\mathrm{min}})}},
\end{equation}
where $\rho$ is the density and $F$ is a constant, roughly equal to
0.34. The quantity $\ln ( {b_\mathrm{max}} / {b_\mathrm{min}})$ is
often denoted by $\ln \Lambda$ and referred to as the Coulomb logarithm.

The appropriate values of $b_\mathrm{min}$ and $b_\mathrm{max}$ in
the above equations have
been discussed at length in the literature. Chandrasekhar (\cite{chandra})
argued that $b_\mathrm{min}$ is the value of $b$ for which the
angular deflection 
of the star is equal to ${1 \over 2} \pi$. The value of $b_\mathrm{max}$
has been subject to considerable controversy. Chandrasekhar (\cite{chandra}),
Kandrup (\cite{kandrup}) and Smith (\cite{smith}) have opted for a
$b_\mathrm{max}$ of 
the order of the mean inter-particle distance, while others
(e.g. Spitzer \& Hart \cite{spitzer2}, Farouki \& Salpeter
\cite{farouki1}, Spitzer \cite{spitzer1}) 
used for $b_\mathrm{max}$ a characteristic radius of the system. The
numerical simulations of Farouki \& Salpeter (\cite{farouki2}) argue
in favour of the latter. This is further corroborated by the
results of Theis (\cite{theis}).

Using the estimates $b_\mathrm{min} = \frac {G m} {\upsilon^2}$,
$b_\mathrm{max}=R$ and 
assuming virial equilibrium, so that we can use for the velocity the
estimate $\upsilon^2 \approx \frac {G N m} {R}$, we  find 
(Binney and Tremaine \cite{BT}) 
\begin{equation}
T_\mathrm{relax}={{N}\over{8~\ln N}}T_\mathrm{cross}.
\label{eq:trel_N}
\end{equation}
Similarly for the case with softening setting $b_\mathrm{min} \ll
\epsilon$, $b_\mathrm{max}=R$ and estimating the velocity by assuming
virial equilibrium we  find 
\begin{eqnarray}
\label{eq:trel_eps}
T_\mathrm{relax} & = & {{\mathrm{\upsilon}^4 R^2}\over{G^2 M^2}}~
{{N}\over {4\left[\ln(R^2/\epsilon^2)-1\right]}}T_\mathrm{cross} \nonumber \\
 & = & {{\mathrm{\upsilon}^4 R^2}\over{G^2 M^2}}~
{{N}\over {8\ln(R/\epsilon)}}T_\mathrm{cross}  \\
 &= &{{N}\over{8\ln(R/\epsilon)}}T_\mathrm{cross} \nonumber
\end{eqnarray}

Equation~(\ref{eq:trel_eps}) differs by a factor of 2 from that
of Huang, Dubinski \& Carlberg 
(\cite{huang}), the reason being that our definition of the relaxation
time is based on the velocities, while that of Huang, Dubinski \&
Carlberg (\cite{huang}) is based on the energies.

It is clear that, for the number of particles used in present day
simulations of collisionless systems and the appropriate values of the
softening, $N$ is considerably larger than $R/\epsilon$. Since,
however, only the logarithms of these quantities enter in equations
(\ref{eq:trel_N}) and 
(\ref{eq:trel_eps}), the differences in the estimates of the relaxation
times differ, for commonly used values of $N$, by less than a factor of
2. Equation (\ref{eq:trel_eps}) 
is more appropriate, since it includes the softening. Often a
coefficient $g$ is introduced in the Coulomb logarithm,
i.e. $\Lambda~=~g N$. Giersz \& Heggie (\cite{Giersz}) estimated
that the most appropriate value of $g$ is 0.11. They also
compiled in their Table 2 the values given by several other
authors. They are all between 0.11 and 0.4. Independent of what is
chosen for the Coulomb logarithm, equations such as (\ref{eq:trel_N})
or (\ref{eq:trel_eps}) argue that even for a moderately low number 
of particles, of the order of say a few thousands, the relaxation time
is comfortably high, of the order of, or higher than, 40 crossing
times. 

\section{Numerical methods}
\label{sec:numerical}

\subsection{Method}

Following Standish \& Aksnes (\cite{standish}), Lecar \&
Cruz-Conz\'alez (\cite{lecar})
and Hernquist (\cite{hernq87}) we will 
measure relaxation using the angular deflection suffered by a
test particle moving in a configuration of $N$ field particles fixed
in space. 

Let us consider a sphere of a given density profile represented by $N$ fixed
particles, which we will hereafter refer to as field particles, and
let us place a test particle of zero mass at the edge of 
this sphere either at rest, or with a radial velocity $v$. In the
limit of $N \rightarrow \infty$ its trajectory
will be a straight line passing through the center of the sphere,
which we will hereafter call the theoretical trajectory. For
a finite $N$, however, the test particle will be
deflected by a number of encounters with the field particles and
thus it will cross the surface of the sphere at an angle $\Phi$ from
the corresponding theoretical point. Following Standish \& Aksnes
(\cite{standish}), or Lecar \& Cruz-Conz\'alez (\cite{lecar}), we can
measure the relaxation time as

\begin{equation}
T_\mathrm{relax}=\frac{T_\mathrm{t}}{sin^2\Phi},
\label{eq:rigid1}
\end{equation}

\noindent
where $T_\mathrm{t}$ is the crossing or transit time.
Different realisations of the adopted model will of course give
somewhat different relaxation times. It is thus better to use an
estimate based on the average of many realisations or
trajectories. Thus we have (Standish \& Aksnes \cite{standish})

\begin{equation}
T_\mathrm{relax}=\frac{<T_\mathrm{t}>}{<sin^2\Phi>},
\label{eq:rigid2}
\end{equation}
where the $<>$ denote an average over all realisations and/or all 
trajectories. 

We will repeat such calculations here, extending them to
non-homogeneous density distributions, 
different initial velocities of the test particle
and a larger range of field particle numbers $N$. This will allow us to
discuss the effect of central concentration, of
initial test particle velocities, of softening and of particle number on the
relaxation time. 

\subsection{Mass distributions and computing miscelanea}
\label{subsec:models}

To find the effect of central concentration on the relaxation time we
use three different 
mass distributions, namely the homogeneous sphere (hereafter model H),
the Plummer model (hereafter model P) and the Dehnen sphere (Dehnen
\cite{dehnen} - hereafter model D). For the Plummer model the density is

\begin{equation}
\rho (r) = \frac {3 M_T} {4\pi a_\mathrm{P}^3} (1 + r^2/a_\mathrm{P}^2)^{-5/2} 
\end{equation}
and for the Dehnen one
\begin{equation}
\rho (r) = {{(3-\gamma) M_T} \over {4 \pi}} {{a_\mathrm{D}} \over
{r^{\gamma}(r+a_\mathrm{D})^{(4-\gamma)}}}, 
\end{equation}

\noindent
where $M_T$ is the total mass of the sphere, $a_\mathrm{P}$ and
$a_\mathrm{D}$ are the two scale-lengths, of the Plummer and Dehnen
models respectively, and $\gamma$ is the concentration index of the latter. 
For the model P we have taken a scale-length of 
$a_\mathrm{P}$ = 9.2, and for model D we have adopted
$a_\mathrm{D}=3.1$ and $\gamma=1$.  In all the examples in
sections~\ref{subsec:npart} to \ref{subsec:g4} we have truncated the mass 
distributions, using a cut-off radius of 20, and taken the mass within this
cut-off radius to be
equal to 15. For these parameters the mass inside the cut-off radius is 
75\% of the total.
For the homogeneous sphere we have adopted a cut-off
radius of 20 and a mass of 15, so as to keep as much in 
line as possible with the other two models. For the gravitational
constant we use $G$~=~1. 

These three models span a large range of central
concentrations. They have the same cut-off radius, but the radius
containing 1/10 of the total mass is for model D roughly one fourth of
the corresponding radius for model P and roughly an order of magnitude
smaller than that of model H. Similarly the radius containing half the
mass for model D is roughly half of that of model P and a third of
that of model H. These models will thus allow us to explore fully the effect of
central concentration on the relaxation time.

We start 1\,000 test particles from random positions on the surface
of a sphere of radius 20 and with initial radial velocities equal to
$f_\mathrm{v}$ times 
their theoretical escape velocity (hereafter $v_\mathrm{esc}$), calculated
from the model. The particles were advanced using a
leap-frog scheme with a fixed time-step of $\Delta t = 2 ^ {-6} =
0.015625$. We made sure this gives a sufficient accuracy by
calculating orbits without the use of GRAPE and with this
time-step. This showed that the energy is 
conserved to 8 or 9 digits. We then repeated the exercise using a
Bulirsch-Stoer scheme (Press et al. \cite{numrec}) and found that the
energy was conserved to 10 digits. Since an accuracy of 8 or 9 digits
is ample for our work, we adopted the leap-frog integrator and the 
afore-mentioned time-step.

The forces between particles were calculated using one of the Marseille
GRAPE-3AF systems, except for the results given in
sections~\ref{subsec:g4} and \ref{subsec:Nbod}, where we used the Marseille
GRAPE-4 system. 

For each model we take 10 different field particles distributions
and for each we calculate the 1\,000 test particles trajectories.
For simplicity the test particles are the same in each of the 10 field
particles distributions.
For each of the test particles we calculate $T_\mathrm{t}$ and 
the deflection angle from the theoretical (straight line)
orbit, $\Phi$. Then the relaxation time is calculated using equations
(\ref{eq:rigid1}) and (\ref{eq:rigid2}). The errors are obtained using
the bootstrap method (Press et al. \cite{numrec}). In Figures
\ref{rigid_N} to \ref{rigid_e} error bars are plotted only when  
$\sigma_{T_\mathrm{relax}}/T_\mathrm{relax} > 0.05$.

\subsection{Validity of the approximations }

\begin{figure}
\resizebox{\hsize}{!}{{\includegraphics*{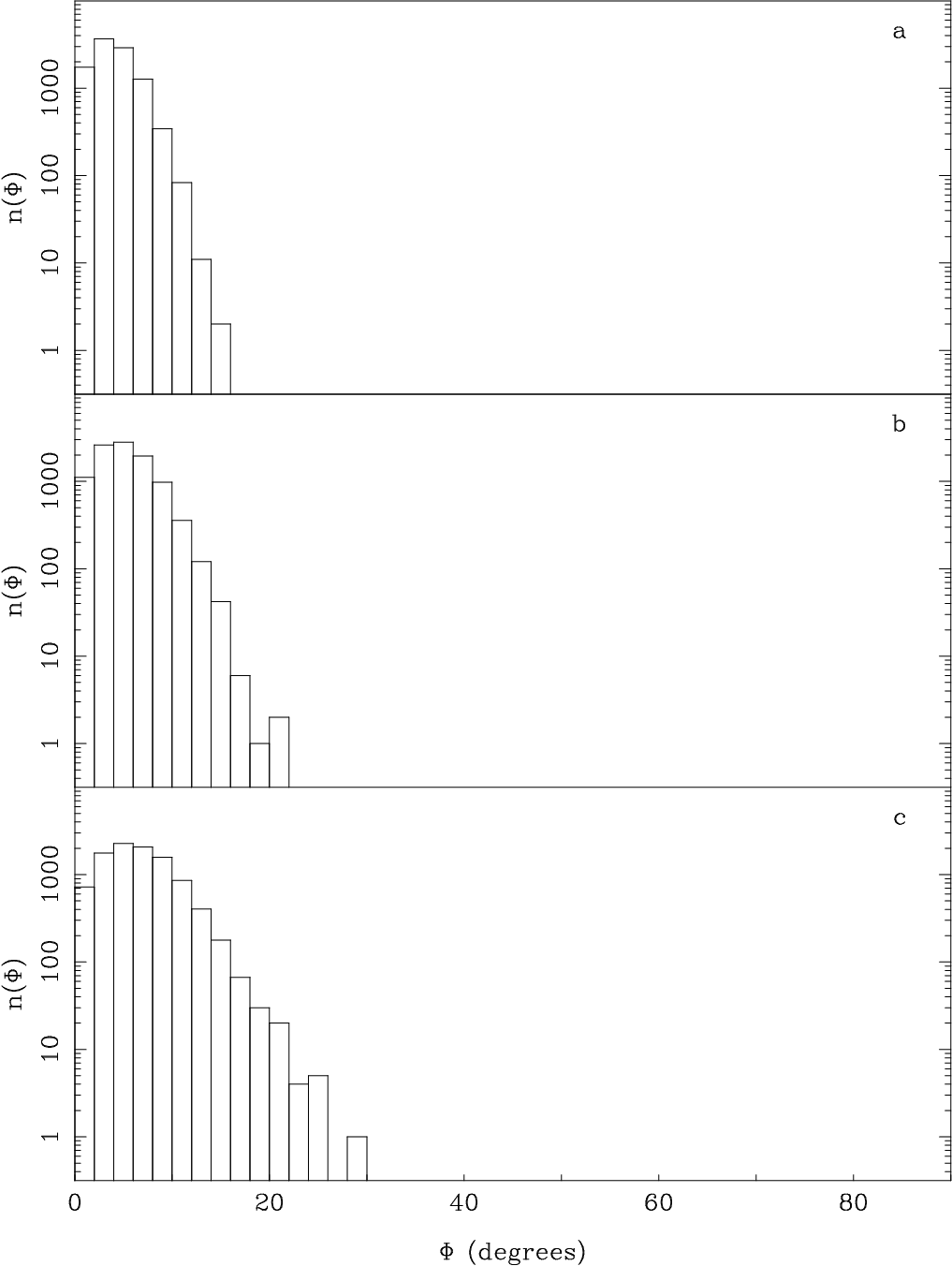}}}
\caption{ 
Distribution of the number of test particles orbits that have a given
deflection angle, $n(\Phi)$, as a function of that angle for model D
and $f_\mathrm{v}=1.5,
\epsilon=0.01$ and $N=4\,000$ (a), $N=2\,000$ (b) or $N=1\,000$ (c).}
\label{fig:defl_angles}
\end{figure}

Equations (\ref{eq:rigid1}) and (\ref{eq:rigid2}) were derived
assuming small deflection angles. There could, however, be
cases in which a test particle comes very near a given field particle
and is very strongly deviated from its initial trajectory, so that the
deflection angle is greater than 
90$\degr$. In that case equations (\ref{eq:rigid1}) and
(\ref{eq:rigid2}) are certainly not valid, particularly since for a
deflection of 180$\degr$ they will give the same result as for
0$\degr$. It is not easy to treat such deflections accurately, so what
we will do here is to keep track of their number and make sure that
it is sufficiently small so as not to influence our results. We have
thus monitored the number of orbits whose velocity component along the
axis which includes the initial radial velocity changes sign. We will
hereafter for brevity call such orbits looping orbits. None
were found for the homogeneous sphere distribution, and very
few, 3 in 10\,000 at the most, for the case of the Plummer sphere,
and that for the smallest of the softenings used here
(cf. section~\ref{subsec:soft} and Figure~\ref{rigid_e}). The largest
number of looping orbits was found, as expected, for model D
and the smallest 
softening, i.e. $\epsilon=2^{-12}$. For this case, $f_\mathrm{v}$ =
1.5 and $N$ = 64\,000, we find of the 
order of 30 such orbits in 10\,000. Although this is considerably
larger than the corresponding number for homogeneous and Plummer
spheres, it is still low enough not to influence much our statistics,
particularly if we take into account that it refers to an exceedingly
small softening. For the more reasonable value $\epsilon=2^{-9}$, we
find that there are no looping orbits at all. We can thus safely
conclude that the number 
of particles with looping orbits is too low to influence our results.

We still have to make sure that the remaining orbits have sufficiently
small deflection angles for equations~(\ref{eq:rigid1}) and
(\ref{eq:rigid2}) to be valid. Figure~\ref{fig:defl_angles} shows a
histogram of the number of test particles orbits that have a given
deflection angle, $n(\Phi)$, as a function of that angle, $\Phi$, for
model D, i.e. the one that should have the largest deflection
angles, with ${f_\mathrm{v}}=1.5$, $\epsilon=0.01$. The upper panel
corresponds to $N=4\,000$, the middle one to $N=2\,000$ and the lower
one to $N=1\,000$. Note that we have used a logarithmic scale for the
ordinate, because otherwise the plot would show in all cases only a few
bins near $0\degr$. 

For $N=1\,000$ there is one particle with deviation of $28\degr$, one
with $26\degr$ and two with $25\degr$. Furthermore only 30
trajectories, of a total of 10\,000, have a deflection angle larger
than $20\degr$. The 
numbers are even more comforting for $N=2\,000$, where only two
trajectories have a deflection angle larger than $20\degr$, and even
more so
for $N=4\,000$, where no particles reach that deflection angle. We can
thus conclude that in all but very few cases the small deflection
angle hypothesis leading to equations~(\ref{eq:rigid1}) and
(\ref{eq:rigid2}) is valid.
 
\section{Free parameters}
\label{sec:free_param}

We will calculate the values of the relaxation time as a function
of three free parameters, namely the number of field particles, the 
initial velocity of the test particles and the softening. In this
section we will discuss what the relevant ranges for these parameters are.

\subsection{Number of particles}
\label{subsec:number}
We will consider numbers of particles between 1\,000 and
64\,000. Indeed fewer than 1\,000 particles are hardly used anymore in
numerical simulations, even with direct summation. 
For more than 64\,000 particles the two-body
relaxation is small. Furthermore the range considered is sufficiently
large for all trends to be clearly seen.

\subsection{Initial velocity}
\label{subsec:velocity}
The question for the initial velocity is more convoluted. The simple
analytical approaches leading to equations~(\ref{eq:trel_N}) and
(\ref{eq:trel_eps}), 
instead of taking a spectrum of velocities for the individual
encounters and then integrating over these velocities, introduce an
effective or average velocity and assume that all interactions are
made at this velocity. In our numerical calculations individual
encounters occur at different velocities, depending on their position
on the trajectory of the test particle and on the initial velocity of
this particle. We can, nevertheless, introduce $v_\mathrm{eff}$, an
average or effective velocity, in a similar way as the analytical
approximation. A simple and straightforward, albeit not
unique, such definition can be obtained as follows:
Let us assume that the test particle moves on a straight line. At each
point of its trajectory we can define a thin sheet going through this
point and locally perpendicular to the trajectory. It will contain all
field particles which have this point as their closest approach with
the test particle. 
Let $dr$ be the  thickness of
this sheet and $\lambda dr$ the fraction of the total mass of the field
particles that is in it. Then we can define the effective velocity

\begin{equation}
v_\mathrm{eff} = 2 \int_0^R \lambda v~ dr,
\label{eq:v_eff}
\end{equation}

\noindent
which of course depends on both the distribution of the field
particles and the initial velocity of the test particle.

\subsection{Softening }
\label{subsec:softening}

\begin{figure}
\resizebox{\hsize}{!}{\rotatebox{-90}{\includegraphics*{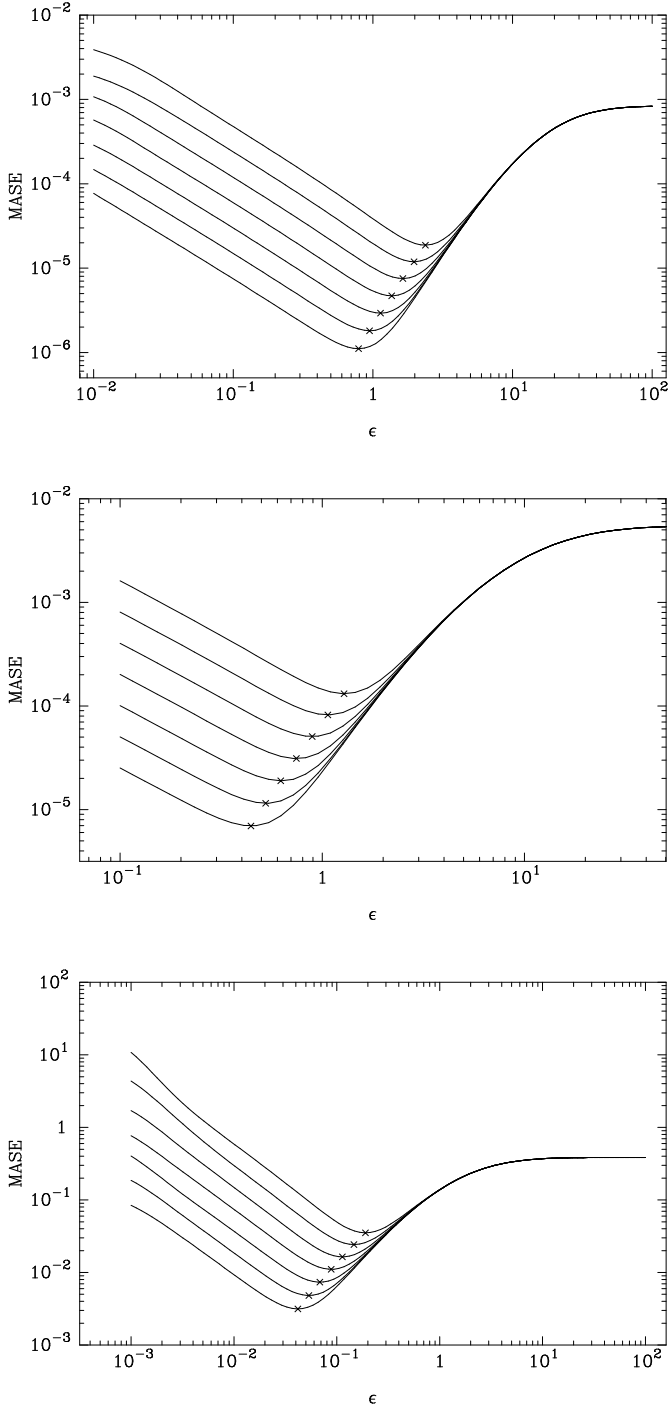}}}
\caption{
$MASE$ as a function of the softening $\epsilon$ for the three models
described in this paper. The upper panel gives the results for model
H, the middle one for model P, and the lower one for model D. In each
panel from top to bottom the curves correspond to $N$ = 1\,000,
2\,000, 4\,000, 8\,000, 16\,000, 32\,000 and 64\,000,
where $N$ is the number of particles in the realisation of each
model. The number of realisations taken in all cases is 6 $\times 10^6 / N$.
The position of the minimum error along a line corresponding to a
given $N$ is marked by a $\times$, and the corresponding $\epsilon$
value is the optimal softening  $\epsilon_{opt}$ for this number of
particles. 
}
\label{fig:mase}
\end{figure}

The third free parameter in our calculations is the softening. Merritt 
(1996; hereafter \cite{M96}) and Athanassoula et al. (2000; hereafter
\cite{AFLB}) 
showed that, for a given mass distribution and a given number 
of particles $N$, there is a value of the softening, called optimal
softening $\epsilon_{opt}$, which gives the most accurate
representation of the gravitational forces within the $N$-body
representation of the mass distribution. For values of the softening
smaller than $\epsilon_{opt}$ the error in 
the force calculation is mainly due to noise, because of the
graininess of the configuration. For values of the softening larger than
$\epsilon_{opt}$ the error is mainly due to the biasing, since the
force is heavily softened and therefore far from Newtonian.
Since the two-body
relaxation is also a result of graininess it makes sense to consider
softening values for which it is the noise and not the bias that
dominates, i.e. to concentrate our calculations mainly on values
of the softening which are smaller than or of the 
order of $\epsilon_{opt}$, keeping in mind that too small values of
the softening have no practical significance. The value of the optimal
softening 
decreases with $N$ and can be well approximated by a power law 

$$\epsilon_{opt}=BN^{b},$$

\noindent
(\cite{M96}). The values of B and b depend on the mass distribution under
consideration, and, to a much smaller degree, on the
range of $N$ considered (\cite{AFLB}). Thus denser 
configurations require smaller softenings for an optimum
representation of the force (\cite{AFLB}). 
For a given number of particles the homogeneous and Plummer
spheres have roughly the same optimal softening, while the Dehnen
sphere with
$\gamma=0$ has an optimal softening 0.45 dex  lower (cf. 
Figure~9 of \cite{AFLB}). 

M96 and \cite{AFLB} have calculated $\epsilon_{opt}$ using the
mean average square error (MASE) of the force, which 
measures how well the forces in an $N$-body representation of a given
mass distribution represent the true forces in this distribution. The average
square error (ASE) is defined as 

\begin{equation}
ASE=\frac{\cal C}{N} \sum_{i=1}^{N}|{\bf F}_i-{\bf
F}_\mathrm{true}({\bf x}_i)|^2, 
\label{eq:ASE}
\end{equation}

\noindent
where ${\bf F}_\mathrm{true} ({\bf x}_i)$ is the true force from
a given mass distribution at
a point ${\bf x}_i$, ${\bf F}_i$ is the force calculated at the same 
position from a given $N$-body realisation of the mass distribution and
using a given  
softening and method, $N$ is the number of particles in the
realisation, and the summation is carried out 
over all the particles. In order to get rid of the dependence on the
particular configuration, which  
is of no physical significance, many realisations of the same 
smooth model must be generated and the results should be averaged over
them. Thus $MASE$, the mean value of the $ASE$, is

\begin{equation}
MASE = \frac {\cal C} {N} <\sum_{i=1}^{N}|{\bf F}_i-{\bf
F}_\mathrm{true}({\bf x}_i)|^2> 
\label{eq:MASE}
\end{equation}

\noindent
where $<>$ indicates an average over many
realisations.  
In equations (\ref{eq:ASE}) and (\ref{eq:MASE}) ${\cal C}$ is a
multiplicative constant, introduced  
to permit comparisons between different mass distributions. Since in
this paper we are only interested in the values of $\epsilon_{opt}$ we
will simply use ${\cal C}$ = 1. The $MASE$ values 
were found using 6 $\times 10^6 / N$ realisations and were calculated
using direct summation on the Marseille GRAPE-5 systems. 

Figure~\ref{fig:mase} shows values of $MASE$ as a function of $\epsilon$ for
the three models considered in subsections~\ref{subsec:npart} to
\ref{subsec:g4} and seven values of $N$, in 
the range of values considered here
(cf. subsection~\ref{subsec:number}). The general form of the curves is
as expected. There is in all cases a minimum error between the
region dominated by the noise (small values of the softening) and the
region dominated by the bias (large values of the softening). This
minimum -- marked with a $\times$ in Figure~\ref{fig:mase} -- gives the
value of $\epsilon_{opt}$. For all three models a larger number of
particles corresponds to a smaller error and a smaller value of
$\epsilon_{opt}$, as expected (\cite{M96}, \cite{AFLB}). 

The more concentrated configurations give smaller values of
$\epsilon_{opt}$, as already discussed in \cite{AFLB}. Thus for $N$ = 64\,000
the optimum softening for model H is less than twice that of model P,
while the ratio between the softenings of models P and D is more than
10.

Comparing our results to those of \cite{AFLB} we can get insight on the
effect of the truncation radius. For this it is best to compare our
results obtained with $N$ = 32\,000 with those given by \cite{AFLB} for $N$ =
30\,000, since these two $N$ values are very close and we do not have
to make corrections for particle number. For our model P
and $N$ = 32\,000 we find an optimum softening of 0.52, or,
equivalently, 0.057$a_\mathrm{P}$, where $a_\mathrm{P}$ is the scale
length of the Plummer sphere. This is smaller than the value
of 0.063$a_\mathrm{P}$ found for the \cite{AFLB} Plummer model and the
difference is  due to the
different truncation radii of the two models. \cite{AFLB} truncated their
Plummer sphere at a radius of 38.71$a_\mathrm{P}$, which encloses
0.999 of the total mass of the untruncated sphere, while model P is
truncated at 2.2$a_\mathrm{P}$, 
which contains only 75\% of total mass. The difference in the values
of $\epsilon_{opt}$ is in good agreement with the discussion in
subsection 5.2 of \cite{AFLB}. When the truncation radius is large, the model
includes a relatively high fraction of low density regions, where the
inter-particle distances are large. This is not the case if the
truncation radius is much smaller, as it is here. Thus the mean
inter-particle distance is larger in the former case and, as can
be seen from Fig. 11 of \cite{AFLB}, the corresponding optimal softening
also. This predicts that the optimal softening should be smaller in
models with smaller truncation radius, and it is indeed what we
find here. 

Our model H is the same as that of \cite{AFLB}, except for the difference in
the cut-off radii. Thus the values of the optimum softening are the
same, after the appropriate rescaling with the cut-off radii has been
applied. Our model D differs in two ways from the corresponding model of
\cite{AFLB}. We use here $\gamma$ = 1, while \cite{AFLB} used $\gamma$
= 0. We also truncated our model at 6.5$a_\mathrm{D}$, while
\cite{AFLB} truncated theirs at 2998$a_\mathrm{D}$. 
It is thus not possible to make any qualitative or quantitative comparisons.

\section{Relaxation measured from angular deflections}
\label{sec:angle}
\indent

\subsection{The effect of the number of particles and of the density
distribution }
\label{subsec:npart}

\begin{table}
\caption[]{Coefficients of linear regression in log-log scale of
$T_\mathrm{relax}$ as a function of particle number $N$}
\begin{flushleft}
\label{tab:rigid_A_B_N}
\begin{tabular}{lllllll}
\hline
Model & $f_\mathrm{v}$ & $\epsilon$ & $A_1$ & $B_1$ & $\sigma_{A_1}$
& $\sigma_{B_1}$ \\ 
\hline\noalign{\smallskip}
 H &     &      & 0.78 & 1.00 & 0.05 & 0.01 \\
 P & 1.5 & 0.01 & 0.76 & 0.98 & 0.04 & 0.01 \\
 D &     &      & 0.63 & 0.78 & 0.04 & 0.01 \\
\noalign{\smallskip}
 H &     &      & 0.87 & 1.01 & 0.05 & 0.01 \\
 P & 1.5 & 0.03 & 0.80 & 1.00 & 0.11 & 0.03 \\
 D &     &      & 0.86 & 0.77 & 0.11 & 0.03 \\
\noalign{\smallskip}
 H &     &      & 0.99 & 1.01 & 0.06 & 0.01 \\
 P & 1.5 & 0.10 & 0.94 & 1.00 & 0.11 & 0.03 \\
 D &     &      & 0.75 & 0.89 & 0.07 & 0.02 \\ 
\noalign{\smallskip}
 H &     &      & 1.13 & 1.02 & 0.08 & 0.02 \\
 P & 1.5 & 0.50 & 1.24 & 0.98 & 0.07 & 0.02 \\
 D &     &      & 1.29 & 0.94 & 0.07 & 0.02 \\ 
\noalign{\smallskip}
 H &     &      & 1.17 & 1.01 & 0.04 & 0.01 \\
 P & 2.0 & 0.03 & 0.98 & 1.01 & 0.08 & 0.02 \\
 D &     &      & 0.78 & 0.87 & 0.12 & 0.03 \\
\noalign{\smallskip}
 P & 3.0 & 0.01 & 1.40 & 0.97 & 0.04 & 0.01 \\
\noalign{\smallskip}
\hline
\end{tabular}
\end{flushleft}
\end{table}

\begin{figure}
\resizebox{\hsize}{!}{{\includegraphics*{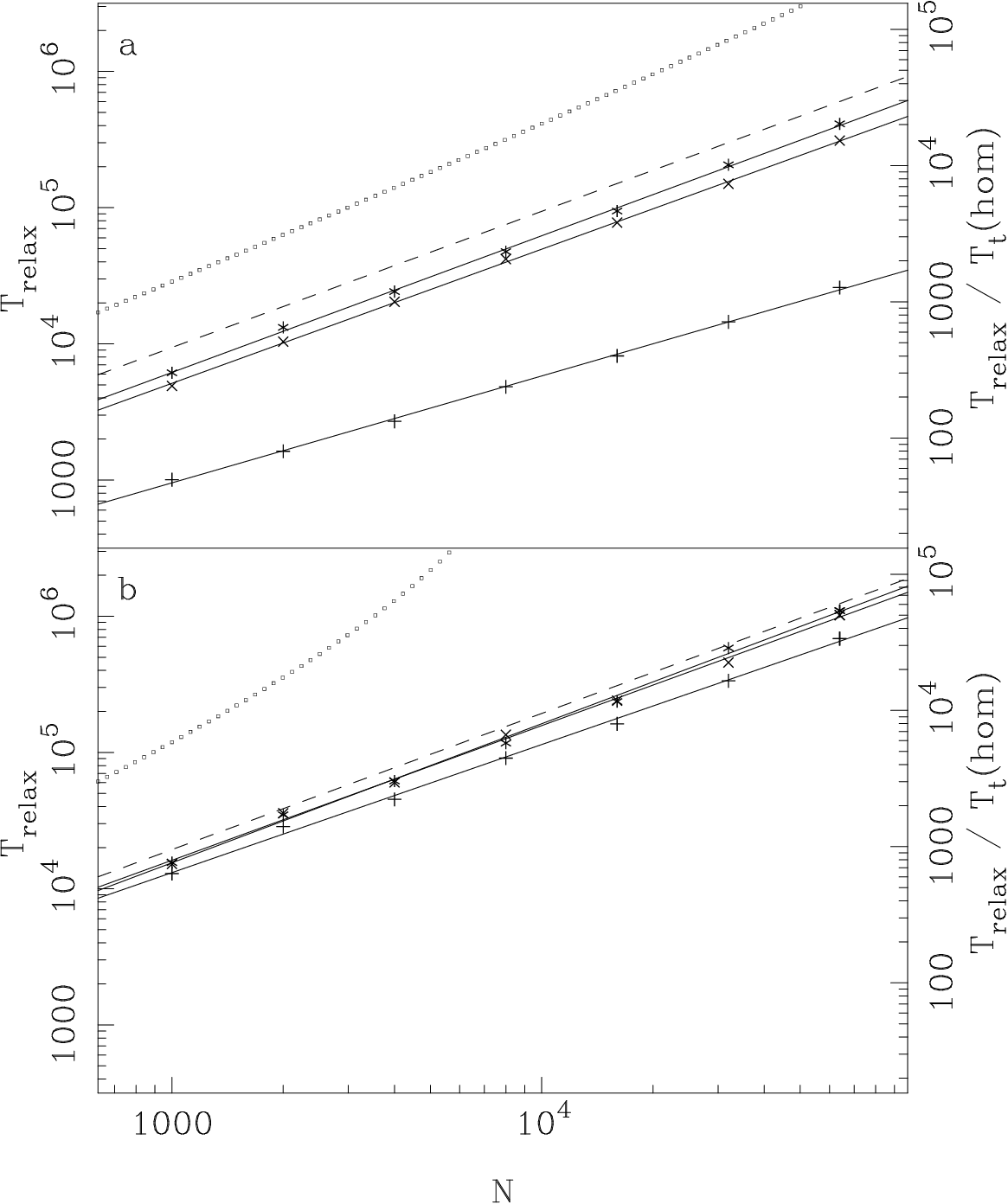}}}
\caption{Relaxation time as a function of the number of field
particles, for $f_\mathrm{v}$=1.5 and three different mass models,
namely model H (stars), model P (x) and model D (crosses). The upper
panel (a) was obtained with $\epsilon=0.01$ and the lower (b) with
$\epsilon=0.5$. The
straight solid lines are the corresponding linear least square
fits. The dashed lines give the predictions of
equation~(\ref{eq:trel_eps}), when we use $b_\mathrm{max}=R$, while the
dotted lines give the prediction of the corresponding equation
obtained by using $b_\mathrm{max}=l$.}
\label{rigid_N}
\end{figure}

Figure~\ref{rigid_N} shows the relaxation time as a function of the number
of particles for the models H, P and D described in the
previous section. The upper panel was
obtained for $f_\mathrm{v}$ = 1.5 and 
$\epsilon$ = 0.01 and the lower one for $f_\mathrm{v}$ = 1.5 and
$\epsilon$ = 0.5. As can be seen from Figure~\ref{fig:mase}, for the
first value of the softening  
noise dominates over bias for all three models. For the second value noise
dominates for model H, bias dominates for model D, and we are near the
optimum value for model P and the high $N$ values. The left ordinate
in fig.~\ref{rigid_N} gives the 
relaxation time while the right one gives the ratio of the relaxation
to transit time for the homogeneous sphere and
$f_\mathrm{v}=1.5$. The corresponding values of this ratio for the
other two density distributions can be easily obtained if one takes
into account that for $f_\mathrm{v}=1.5$ the three theoretical
transit times are 20.35, 17.52 and 16.91, for the H, P and D
models respectively. 
We fitted straight lines 
$$\log_{10}(T_\mathrm{relax}) = A_1 + B_1 \log_{10}(N)$$ 
to 
each model using least-square 
fits and plotted them with solid lines in the figure. The values of
$A_1$ and $B_1$, as well as of their corresponding 
uncertainties $\sigma_{A_1}$ and $\sigma_{B_1}$, are given in
Table~\ref{tab:rigid_A_B_N}. We have  
performed similar calculations for other values of the velocity and
softening parameters and
have included in this Table the corresponding values of $A_1$ and $B_1$,
as well as their uncertainties.
From Figure~\ref{rigid_N}, other similar plots for other values
of the softening and of the initial velocity (not presented here), and
the values given in Table~\ref{tab:rigid_A_B_N} we can reach a number
of conclusions. 

The dependence of $T_\mathrm{relax}$ on $N$ is reasonably well represented by a
straight line in the log-log plane and that for all values of the
softening and $f_\mathrm{v}$ we tried and for all three models. The
relaxation time for model D 
is always smaller than that for models P and H.
The difference is much more important ($\sim$ 1 dex) for the smaller
value of the softening, than for the larger one 
($\sim$ 0.15 dex). Similarly, the relaxation time for Plummer
distributions is somewhat smaller ($\sim$ 0.1 dex) than that of the homogeneous
one for small values of the softening, while for the larger value the
two relaxation times do not differ significantly. 

Figure~\ref{rigid_N} and Table~\ref{tab:rigid_A_B_N} also show a trend
for the slopes of the straight lines. For the small softening the
homogeneous model has a relaxation time which is roughly proportional
to the number of 
field particles in the configuration, and that is in good agreement 
with equation~(\ref{eq:trel_eps}). This is not true any more for the
more centrally concentrated mass distributions, which have a value of $B_1$
less than unity. For model P the value of  $B_1$ is slightly less than
unity, but for model D it is considerably so. Thus when
we increase the number of particles in the configuration, we increase
the relaxation time relatively less in more centrally concentrated
configurations than in less centrally concentrated ones. In other
words not only is the relaxation time 
smaller in the more centrally concentrated configurations, but also
it takes more particles to increase it by a given amount. 

These trends for the slopes of the straight lines are also present for
the larger value of the softening. The
differences, however, between the slopes, i.e. between the
corresponding values of $B_1$, are much
smaller. Thus for $f_\mathrm{v}=1.5$ and $\epsilon=0.01$  there is a
difference of roughly 25\% 
between the slopes corresponding to models H and D, while this value is
reduced to roughly 8\% when $\epsilon=0.5$.

Figure~\ref{rigid_N} also compares the prediction of
equation~(\ref{eq:trel_eps}) 
with the results of our numerical calculations for the homogeneous
sphere. The agreement is fairly good, particularly for the larger
softening, where the difference is of the order of 0.08 dex.
It should be noted that these predictions were obtained with
$b_\mathrm{max}=R$, 
the cutoff radius of the system. A yet better agreement would have
been possible if one used $b_\mathrm{max}=fR$, where $f$, a constant
larger than 1. Since, however, this constant is a function of the
softening used and perhaps also of the value of $f_\mathrm{v}$,
it is not very useful to determine its numerical value. The results obtained by
using $b_\mathrm{max}=l$, where $l$ is the mean inter-particle distance, are
given by open squares. We note that they give a bad approximation of
the numerical results, particularly so for a large number of particles
and for the value of the softening
which is nearest to optimal. For the smaller softening the
approximation with $b_\mathrm{max}=l$ is still considerably worse than
that obtained with 
$b_\mathrm{max}=R$, but the difference is smaller than in the case of
the optimal softening. Whether this 
will be reversed for an even smaller value of the softening or not is
not possible to predict from the above calculations. Nevertheless, if
it did happen, it would be for a value of the softening that gave a
very bad representation of the forces within model H.
Thus we can conclude that, at least for collisionless
simulations which have a reasonable softening, the simple analytical estimates
presented in section~\ref{sec:analytical} give a
reasonable approximation of the relaxation time if we use
$b_\mathrm{max}=R$, but not if we use $b_\mathrm{max}=l$. The latter
gives too high a value of the relaxation time, and is therefore
falsely reassuring. 

We also compared our results with theoretical estimates using
$\ln\Lambda~=~\ln(gN)$ for the Coulomb logarithm and different
values of $g$, as tabulated by Giersz \& Heggie (\cite{Giersz}). We
find that they always fare less well than equation~(\ref{eq:trel_eps}) with 
$b_\mathrm{max}=R$, particularly for $\epsilon=0.01$. Amongst the
values tried, 
$g$~=~ 0.11, proposed by Giersz \& Heggie (\cite{Giersz}),
gave the best fit for $\epsilon=0.5$, while the value $g$~=~ 0.4
(Spitzer \cite{spitzer1}) was best for $\epsilon=0.01$. The
differences between the results for various values of $g$ are
nevertheless small.  

To summarise this section we can say that more centrally 
concentrated distributions have smaller relaxation
times and that the difference is more important for smaller values of
the softening. This argues that the relaxation
time is more influenced by the maximum rather than by the average
density. 

\subsection{The effect of velocity}
\label{subsec:vel}

\begin{figure}
\resizebox{\hsize}{!}{{\includegraphics*{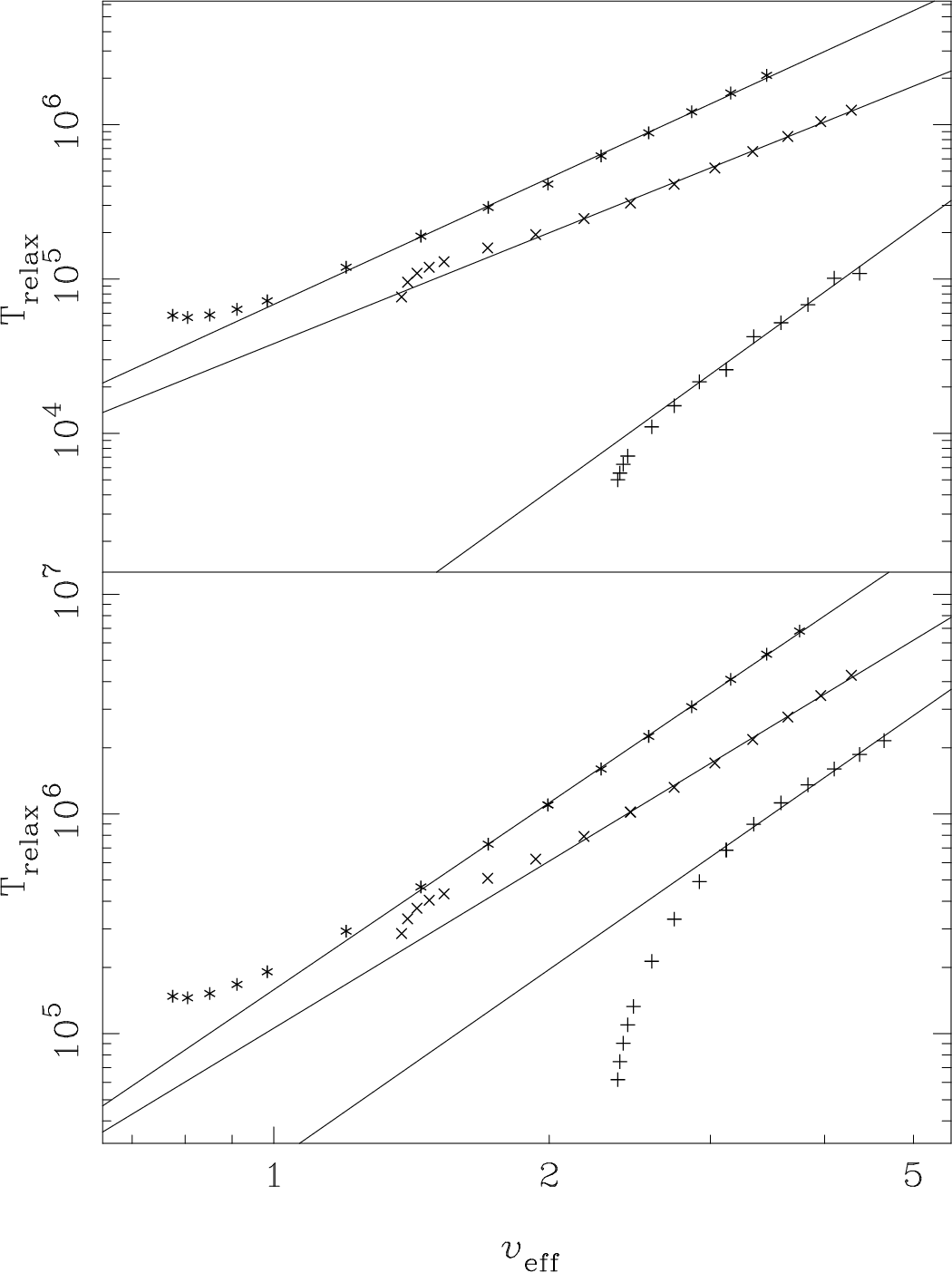}
}}
\caption{Relaxation time as a function of the effective velocity
of the test 
particles, for $N$=64~000 and three different mass models,
namely model H (stars), model P (x) and model D (crosses). The
straight lines are the corresponding linear least square fits. The
upper panel corresponds to $\epsilon$=0.01 and the lower one to 
$\epsilon$=0.5 }
\label{rigid_v}
\end{figure}

\begin{table}
\centering
\caption{Coefficients of linear regression in log-log scale of
$T_\mathrm{relax}$ as a function of $v_\mathrm{eff}$}
\begin{flushleft}
\label{tab:rigid_A_B_v}
\begin{tabular}{llllll}
\hline
Model & $\epsilon$ & $A_2$ & $B_2$ & $\sigma_{A_2}$ & $\sigma_{B_2}$ \\
\hline
 H &      & 4.84 & 2.72 & 0.02 & 0.04 \\ 
 P & 0.01 & 4.58 & 2.39 & 0.02 & 0.04 \\ 
 D &      & 2.33 & 4.28 & 0.20 & 0.35 \\
\noalign{\smallskip}
 H &      & 5.20 & 2.83 & 0.01 & 0.02 \\ 
 P & 0.5  & 5.02 & 2.53 & 0.02 & 0.04 \\ 
 D &      & 4.42 & 2.90 & 0.06 & 0.11 \\
\hline
\end{tabular}
\end{flushleft}
\end{table}

In order to test
the effect of velocity on the relaxation time we have launched test
particles with different initial velocities.

Figure~\ref{rigid_v} shows the relaxation time as a function of the
effective velocity of the particles, defined in
section~\ref{sec:free_param}, for the
three density 
distributions under consideration. The calculations have been made
with 64~000 field particles and a softening of 0.01 for the upper
panel and 0.5 for the lower one. The dependence is linear in the
log-log plane for large values of the effective velocity and deviates
strongly from it for small values. We thus fitted a
straight line in the log-log plane to the higher velocities estimating
for each of the mass models separately the number of points that could
be reasonably fitted by a straight line. We give the constants of the
regression 

$$\log_{10}(T_\mathrm{relax}) = A_2 + B_2 \log_{10} ({v_\mathrm{eff}}),$$

\noindent
together with the corresponding error estimates, in
Table~\ref{tab:rigid_A_B_v}. 

In all cases the relaxation
time increases with the initial velocity of the particles. In order to
compare this with the analytical predictions of
section~\ref{sec:analytical} we note that the deviation of a particle
from its trajectory due to an encounter 
should be smaller for larger impact velocities (cf. equation
\ref{eq:single_dev_BT}), while larger 
impact velocities imply smaller crossing times. Thus from  equations
(\ref{eq:trel_N}) and (\ref{eq:trel_eps}) we note that 
$T_\mathrm{relax}$ should be proportional to the third power of the
velocity, i.e. that the coefficient $B_2$ in Table~\ref{tab:rigid_A_B_v}
should be roughly equal to 3. We note that in the homogeneous case,
which should be nearer to the analytical result, there is a difference
of less than 10\%, presumably due to the fact that the 
approximations of the analytical approach are too harsh. The
differences with the results of models P and D are on average larger, but
strictly speaking, equations (\ref{eq:trel_N}) and (\ref{eq:trel_eps})
do not apply to them. 

\subsection{The effect of softening}
\label{subsec:soft}

\begin{figure}
\resizebox{\hsize}{!}{\rotatebox{-90}{\includegraphics*{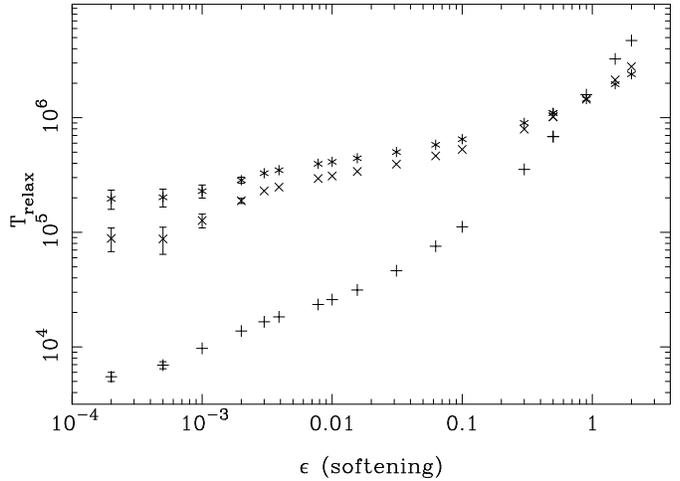}}}
\caption{Relaxation time as a function of the softening,
for $N$=64~000, $f_\mathrm{v}$=1.5 and three different mass models,
namely model H (stars), model P (x) and model D (crosses).}
\label{rigid_e}
\end{figure}

Figure \ref{rigid_e} shows the relaxation time $T_\mathrm{relax}$ as a function
of the softening $\epsilon$ for the case of $N$~=~64,000  and
$f_\mathrm{v}$=1.5. We note that the relaxation time increases with
softening as expected. There is no simple linear
dependence between the two plotted quantities. In fact the relaxation
time increases much faster with $log~\epsilon$ for large values of the
softening than for small ones. The point at which this change of slope
occurs is roughly the same for models H and P, 
and much smaller for model D. In fact in all cases
it is roughly at the position of the corresponding optimal softening,
which is roughly the same for models H and P and
considerably smaller for model D (cf. section
\ref{subsec:softening}). Thus the change of 
slope must correspond to a change between a noise dominated regime and
a bias dominated one. 

In the noise dominated regime the sequence between the three models is
the same as in previous 
cases. Namely it is model H that has the largest
relaxation time, followed very closely by model P, and less
closely by model D. It is interesting to note that
for high values of the softening the three curves intersect. Such
values, however, are too dominated by bias to be relevant to $N$-body
simulations. 

Figure \ref{rigid_e} shows the only examples in this paper in
which the error bars are large enough to be 
plotted, i.e. for which $\sigma_{T_\mathrm{relax}}/T_\mathrm{relax} > 0.05$. These
occur for the smallest values of the softening,  
used here more for reasons of completeness than for their practical
significance. 

\subsection{GRAPE-3 results compared to GRAPE-4 results }
\label{subsec:g4}

All results presented so far were made on one of the Marseille GRAPE-3
systems (Athanassoula et al. \cite{ABLM98}). Such systems, however,
are known to have limited 
accuracy, since they use 14 bits for the masses, 20 bits for the
positions and 56 bits for the forces. One could thus worry whether
this would introduce any extra graininess, which in turn would decrease the
relaxation time.

\begin{table}
\caption[]{Relaxation time $T_\mathrm{relax}$ of model D with
$f_\mathrm{v}=1.5$ and $\epsilon=0.01$, obtained by the GRAPE-3 (G3)
and GRAPE-4 (G4) machines.}
\begin{flushleft}
\label{tab:rigid_G3_G4}
\begin{tabular}{cccc}
\hline
N & G3 & G4 & $\|$G3-G4$\|$/G4 \\
\hline\noalign{\smallskip}
 1\,000  &  1\,040   &   1\,007   &  0.0328  \\
 2\,000  &  1\,626   &   1\,633   &  0.0043  \\
 4\,000  &  2\,698   &   2\,707   &  0.0033  \\
 8\,000  &  4\,834   &   4\,845   &  0.0023  \\
16\,000  &  8\,143   &   8\,123   &  0.0025  \\
32\,000  & 14\,490   &  14\,548   &  0.0040  \\
64\,000  & 25\,858   &  25\,860   &  0.0001  \\
\noalign{\smallskip}
\hline
\end{tabular}
\end{flushleft}
\end{table}

In order to test this we repeated on GRAPE-4, which is a high accuracy
machine, some of the calculations made also with GRAPE-3. For this
purpose, we ran 7 configurations of model D with
$f_\mathrm{v}=1.5$, $\epsilon=0.01$ and different number of field
particles. The calculated relaxation time, $T_\mathrm{relax}$,
obtained by the GRAPE-3 and GRAPE-4 runs, together with their
differences, are shown in Table~\ref{tab:rigid_G3_G4}. As we can see
the results have, in all but one case, a difference less than 0.5\%.
Only in the case of $N=1\,000$ does the difference rise to 3\%, but as we
mentioned before, this very low number of particles is hardly used 
nowadays, even in direct summation simulations on a general purpose computer.

\subsection{Predicting the value of $T_\mathrm{relax}$ for an $N$-body simulation }
\label{subsec:Nbod}

In the standard version of the angular deflection method we have used
so far all the test  
particles start from the same radius with the same initial velocity. On the
other hand in any $N$-body realisation of a 
given model the particles have different apocenters. It is thus
necessary to extend this method somewhat in order to obtain
an estimate of how long a given $N$-body
simulation will remain unaffected by two-body relaxation. Let us
consider a simple model consisting of a Plummer sphere of total mass
20 and scale length 9.2, truncated at 
a radius equal to 30.125, i.e. at a radius containing roughly 7/8 of the total
mass. As an example we wish to estimate the relaxation time of a
74668-body\footnote{This value of $N$ was chosen in order to have
64\,000 field particles within a radius of 20}
realisation of this model which will be evolved with 
a softening of 0.5, a value near the optimal softening for
model P. For this we will somewhat modify the standard angular
deflection method in order to
consider several groups, starting each at a given radius. We first calculate
the relaxation time from each group separately. The relaxation time of
the model will be a weighted average of the relaxation times of the
individual groups. The weights have to be calculated in such a way
that the mean velocities with which the test particles encounter the
field particles at any given radius represent fairly well the encounter
velocities between any two particles in the $N$-body realisation,
which is not far from the dispersion of velocities. We found we could
achieve this reasonably well by considering 18 groups, of 10\,000 test
particles each, with apocenters $R_{max} (i) = 1.25 i + 2.5, i =
1,..18$. Each group starts from a distance such that $f_\mathrm{v} =
0.2$ and 
we weight the results of each by $3^{-i}, i = 1,..18$. These weighting factors
were just found empirically after a few trials and deemed adequate
since they give 
an approximation of the velocity dispersion of the Plummer sphere of
the order of 10 per cent. It would of course be
possible to get a better representation by using a larger number of
groups and e.g. some linear programming technique, but since we only 
need to have a rough approximation of the 
encounter velocities and the fit we obtained is adequate, we did not
deem it necessary to complicate the problem unnecessarily. 

As expected, we find that the relaxation time and the transit time are 
larger for groups with larger initial radius. The range of values we
find is rather large. Thus for the innermost group we find a
relaxation time of 1.6 $\times 10^4$ and a transit time of 4, while
for the outermost group the corresponding values are 3.6 $\times 10^5$
and 51. The  
weighted average of the relaxation time, taken over all orbits in 
the representation, is 3.4 $\times 10^4$, and that of the transit
time 4.9, i.e. nearer to those of the inner groups due to their larger
weights. 

A comparison with the estimates of the simple analytical formula given
in equation (7) is not straightforward, since one has to adopt a
characteristic radius, and the result is heavily dependent on
that. Thus if we adopt an outer radius, where the virial velocity is
small and therefore the crossing time large, we obtain very large
values of the relaxation time, like those we find for the outer parts
of our model Plummer sphere. On the other hand if we take the half
mass radius then we obtain $T_\mathrm{relax}=3.1 \times 10^4$, in good
agreement with our estimate obtained from the weighted average of all
groups. 

Our model P is sufficiently similar to the
$W_c = 5$ King model used by Huang, Carlberg \& Dubinski (\cite{huang}) to
allow comparisons. These authors obtain the relaxation time by
monitoring the change of energy of 
individual particles in a simulation. They consider only particles which at
the end of the 
simulation are near the half mass radius and find a
relaxation time which, rescaled to our units, is 2.5 $\times
10^5$. Thus this estimate is based on a
group of particles which have their apocenters at or beyond the half mass
radius. Applying our own method only to such particles also we
obtain a relaxation time of 2.3 $\times 10^5$, which is in excellent agreement
with the value of Huang et al.(\cite{huang}).

\section{Summary and discussion}
\label{sec:discussion}
\indent

In this paper we have calculated two-body relaxation times for
different mass distributions, number of particles, softenings and
particle velocities. For this we launched
test particles in a configuration of rigid field particles and measured
the relaxation time from the deflection angles (measured from the
theoretical trajectory of the same particles) and the transit
times.  

We first determine the range of softening values for which the error
in the force calculation is dominated by noise, rather than by
bias. These extend to larger values of the softening for smaller
number of particles and for less centrally concentrated
configurations, in good agreement with what was found by \cite{AFLB}. We also
find them to be somewhat larger for models with a larger cut-off radius.

We confirm that the relaxation time increases with the number of
particles. Indeed a larger number of particles entails a lesser
graininess and thus a smaller effect of two-body encounters. In
particular for homogeneous density distributions we confirm the
analytical result that the relaxation time is proportional
to the number of particles. We find, however, that this
dependence does not hold for all mass distributions. 

We find that the relaxation time depends strongly on how centrally
concentrated the mass distribution is, in the sense that more
centrally concentrated configurations have considerably shorter
relaxation times. This can be understood if we make an $N$-body
realisation not by distributing particles of equal masses in such a
way as to follow the density, but by distributing the particles
homogeneously in space and attributing to each one of them an
appropriate mass. Our effect then simply follows
from the fact that the deviation of a
particle trajectory by a single very massive particle is larger
than that due to a sum of low mass particles of the same total mass.

This dependence of relaxation time on central concentration is
strong. E.g. for a softening  
$\epsilon=0.01$ and a $f_\mathrm{v}$ of 1.5 the relaxation times of
models D and H (or P) differ by roughly an order of magnitude. In order to
achieve this difference by changing the number of particles one has to
increase them by a factor of 10 also. I.e. in order to avoid two-body
relaxation ten times more particles are necessary for a simulation
of model D than for one of model H, provided one
uses the same softening. The difference is even larger if the
softening is chosen to be optimal in each configuration, since the 
optimal softenings differ by more than an order of magnitude, so an
extra factor of two is introduced to the necessary particle number.

Also the dependence of the relaxation time on the number of particles
changes with the central concentration of the configuration. We find a
shallower dependence for our more concentrated models, the difference
being more important for smaller values of the softening. For a
softening of 0.01 the difference in the exponent of the power law
dependence is of the order of 20\%.

We find that the relaxation time increases with velocity, as
expected. The reason is that the deflections in two-body encounters
are larger when the relative velocity is smaller. The dependence of
the relaxation time on the effective velocity is linear in a log-log
plane for the larger values of the effective velocity we have
considered and deviates
strongly from linear for smaller velocities. In the simple analytical
estimates of 
section~\ref{sec:analytical}, $T_\mathrm{relax}$ is proportional to the third
power of the velocity. Our more precise numerical estimates argue that
these estimates are only about 10 \% off for the case of the
homogeneous sphere, to which they apply.

The strong decrease of relaxation time with encounter velocities entails
that two-body relaxation has little effect in simulations of
``violent'' events as 
collapses or mergings. On the other hand it may, depending on the
configuration, the number of particles and the softening, play a role
in simulations of quasi-equilibrium configurations. Furthermore
two-body relaxation will be less of a menace in simulations of
objects with high velocity dispersions, like giant ellipticals which
are largely pressure supported, than in cases with less pressure
and more rotational support, like small ellipticals or discs,
putting aside of course 
the effects of shape and rotation, which we have not addressed here.

We have also examined the dependence of the relaxation time on 
softening. We find that, as expected, the relaxation time increases
with softening. The dependence, however, is complicated, and not
given by a simple mathematical formula. Nevertheless for the not too
centrally concentrated models the increase is not too large. Thus for
our models H and P we increase the relaxation time by a factor of the
order of 2 if we increase the softening by a factor of 10. In this we
agree well with Theis (\cite{theis}). The only case where the increase is more
pronounced is for model D and particularly for high values of the
softening. It should, 
however, be remembered that this is a bias dominated regime. We can
thus conclude that in the noise dominated regime the increase of the
relaxation time with the softening is relatively small.

Finally we compared results obtained using GRAPE-3 with those found by
GRAPE-4, and found they are similar. From the above results we can
deduce that the relaxation times  
of the two types of GRAPE systems are essentially the
same. This can be understood as follows. Athanassoula et
al. (\cite{ABLM98}) argued that the limited precision of GRAPE-3 does
not influence the 
accuracy with which the force is calculated since the error in the
calculation of the pairwise forces  can be
considered as random (cf. their Figure 5). This is in good agreement
with what was initially argued
by Hernquist, Hut \& Makino (\cite{hernq93}) and Makino
(\cite{makino1}), who pointed out 
that the two-body relaxation
dominates the error and that the effect of the error in the force
calculation is practically negligible, provided this error is random.

Our results for the relaxation time are always smaller than those
given by the simple analytical formula. The differences are relatively
small if one uses $b_\mathrm{max}=R$ in the formula, but quite large
if one uses $b_\mathrm{max}=l$. Thus our results argue strongly that
the former is the right value to use, at least for collisionless
simulations. In this we agree with Spitzer \& Hart (\cite{spitzer2}),
Farouki \& Salpeter (\cite{farouki1}), Spitzer (\cite{spitzer1}) and
Theis (\cite{theis}). It should also be 
noted that the analytical estimate obtained with $b_\mathrm{max}=l$ is
always considerably larger that the numerical result, and thus is falsely 
reassuring.

By extending somewhat the standard method based on the angular
deflections we obtained an estimate for the relaxation time of an
$N$-body simulation of a Plummer sphere. Comparing it with the results
found by Huang et al. (\cite{huang}) we find excellent agreement. This
is very interesting since the method used by Huang et
al. (\cite{huang}) obtain their estimate of the relaxation time
directly from an $N$-body simulation, i.e. include collective 
effects. This agreement could argue that such effects are not very
large, and thus 
gives more weight to the results obtained with our simple and
straightforward method. 

It is often stressed that galaxies have so many stars that their
relaxation times are far longer than the age of the universe, and thus
that $N$-body simulations extending to long periods of time should
have a very large number of 
particles to also achieve sufficiently large relaxation times. As a
counter-argument one could say that real galaxies are not only composed of
individual stars, but also of star clusters and gaseous clouds, which,
being considerably more massive than individual stars, will introduce
two-body relaxation and change the dynamics from that of a
collisionless system. This, however, is no argument in favour of
$N$-body simulations with short relaxation times, since the deviations
from the evolution of a smooth stellar fluid brought by the graininess
of the $N$-body system need not be the same as those brought by the
compact objects, star clusters or gaseous clouds. It is thus necessary
in $N$-body simulations to strive for high relaxation times 
and believe the results only for times considerably
shorter than that. If desired, the effects of the compact objects,
star clusters or gaseous clouds can then be studied separately. 

\begin{acknowledgements}
We would like to thank Albert Bosma for
many useful discussions.
We would also like to thank IGRAP, the
INSU/CNRS, the region PACA and the University of Aix-Marseille I for
funds to develop 
the computing facilities used for the calculations in this paper.
C.V. acknowledges the European Commission ERBFMBI-CT95-0384
T.M.R. postdoctoral grant and
the hospitality of the Observatoire de Marseille. 
\end{acknowledgements}

\end{document}